\documentclass[aps,prc,twocolumn,showpacs,floatfix,nofootinbib]{revtex4}

\usepackage{amsmath,amssymb,amsfonts,graphicx,yhmath}
\usepackage[colorlinks=true,linktocpage=true,linkcolor=blue,citecolor=blue]{hyperref}
\usepackage[usenames,dvipsnames]{color}
\usepackage{float}

\usepackage{soul} 

\begin{document}

\preprint{}

\title{A viscous blast-wave model for relativistic heavy-ion collisions}

\author{Amaresh Jaiswal}
\affiliation{GSI, Helmholtzzentrum f\"ur Schwerionenforschung, Planckstrasse 1, D-64291 Darmstadt, Germany}
\author{Volker Koch}
\affiliation{Lawrence Berkeley National Laboratory, Nuclear Science Division, MS 70R0319, Berkeley, California 94720, USA}

\date{\today}

\begin{abstract}

Using a viscosity-based survival scale for geometrical perturbations 
formed in the early stages of relativistic heavy-ion collisions, we 
model the radial flow velocity during freeze-out. Subsequently, we 
employ the Cooper-Frye freeze-out prescription, with first-order 
viscous corrections to the distribution function, to obtain the 
transverse momentum distribution of particle yields and flow 
harmonics. For initial eccentricities, we use the results of Monte 
Carlo Glauber model. We fix the blast-wave model parameters by 
fitting the transverse momentum spectra of identified particles at 
the Large Hadron Collider (LHC) and demonstrate that this leads to a 
fairly good agreement with transverse momentum distribution of 
elliptic and triangular flow for various centralities. Within this 
viscous blast-wave model, we estimate the shear viscosity to entropy 
density ratio $\eta/s\simeq 0.24$ at the LHC.

\end{abstract}



\maketitle

\section{Introduction}

High-energy heavy-ion collision experiments at the Relativistic 
Heavy Ion Collider (RHIC) \cite{Adams:2005dq, Adcox:2004mh} and the 
Large Hadron Collider (LHC) \cite{ALICE:2011ab, ATLAS:2012at, 
Chatrchyan:2013kba} have conclusively established the formation of a 
strongly interacting Quark-Gluon Plasma (QGP). The QGP formed in 
these collisions exhibit strong collective behaviour, and therefore 
can be studied within the framework of relativistic hydrodynamics. 
The hydrodynamical analyses of the flow data suggests that the QGP 
behaves like a nearly perfect fluid with an extremely small shear 
viscosity to entropy density ratio $\eta/s$ \cite{Romatschke:2007mq, 
Song:2007ux}. Local pressure gradients due to deformations and 
inhomogeneities in the the initial stages of the collision results 
in anisotropic fluid velocity. These anisotropies subsequently 
translate into flow harmonics describing the momentum asymmetry of 
produced particles \cite{Heinz:2013th, Gale:2013da}. The $\eta/s$ of 
the fluid governs the conversion efficiency, which results in a 
suppression of elliptic flow and higher order flow coefficients \cite
{Luzum:2008cw, Song:2007fn, Dusling:2007gi, Steinheimer:2007iy, 
Molnar:2008xj, Bozek:2009dw, Chaudhuri:2009hj, Xu:2007jv, 
Schenke:2010rr, Holopainen:2010gz, Qiu:2011iv, Qiu:2011hf, 
Song:2010mg, Bhalerao:2015iya}.

Apart from hydrodynamics, the observables pertaining to collective 
behaviour of QGP can also be studied using the so-called blast-wave 
model. Using a simple functional form for the phase-space density at 
kinetic freezeout, Schnedermann {\it et. al.} \cite 
{Schnedermann:1993ws} approximated hydrodynamical results with 
boost-invariant longitudinal flow \cite{Bjorken:1982qr}. They used 
this blast-wave model to successfully fit the transverse momentum 
spectra with only two parameters: a kinetic temperature, and a 
radial flow strength. However, this model was only valid for central 
collisions at midrapidity. In order to make it applicable for 
non-central collisions, Huovinen {\it et. al.} \cite 
{Huovinen:2001cy} generalized this model to account for the 
anisotropies in the transverse flow profile by introducing an 
additional parameter. This new parameter controlled the difference 
between the strength of the flow in and out of the reaction plane. 
This lead to a fairly good fit with the measured elliptical flow as 
a function of transverse momentum. However, the STAR Collaboration 
achieved better fits when they generalized the model even further by 
introducing a fourth parameter to account for the anisotropic shape 
of the source in coordinate space \cite{Adler:2001nb}. Teaney made 
the first attempt to estimate the effect of viscosity on elliptic 
flow using a variant of the blast-wave model model \cite 
{Teaney:2003kp}. However, the centrality dependence of the fit 
parameters left the model with little predictive power.

In this paper we generalize the blast-wave model to include viscous 
effects by employing a viscosity-based survival scale for 
geometrical anisotropies, formed in the early stages of relativistic 
heavy-ion collisions, in the parametrization of the radial flow 
velocity. The present model has five parameters, including $\eta/s$, 
which has to be fitted only for one centrality. In the Cooper-Frye 
freeze-out prescription for particle production, we consider the 
first-order viscous corrections to the distribution function \cite
{Teaney:2003kp}. In essence, we provide a model which incorporates 
the important features of viscous hydrodynamic evolution but does 
not require to do the actual evolution. We use this viscous 
blast-wave model to obtain the transverse momentum distribution of 
particle yields and flow harmonics for LHC. The blast-wave model 
parameters are fixed by fitting the transverse momentum spectra of 
identified particles. Subsequently, we show that this leads to 
fairly good agreement with transverse momentum distribution of 
elliptic and triangular flow for various centralities as well as 
centrality distribution for integrated flow. We estimate the shear 
viscosity to entropy density ratio $\eta/s\simeq 0.24$ at the LHC, 
within the present model.


\section{Blast wave model}

The blast-wave model has been used extensively to fit experimental 
data and it provides good description of spectra and elliptic flow 
observed in relativistic heavy-ion collisions \cite{Huovinen:2001cy, 
Adler:2001nb, Teaney:2003kp, Tang:2008ud, Sun:2014rda}. The 
previously used blast-wave models employ a simple parametrization 
for the flow velocity of boost invariant ideal hydrodynamics. The 
most important feature is the parametrization of the transverse 
velocity which is assumed to increase linearly with respect to the 
radius. This parametrization is found to be in agreement with hydro 
results \cite{Teaney:2001av, Habich:2014jna}. This essentially leads 
to an exponential expansion of the fireball in the transverse 
direction, hence the term blast-wave. Apart from boost invariance, 
the model also assumes rotational invariance. In the following, we 
quickly outline the key features of the blast-wave model.

In order to consider a boost invariant framework, it is easier to 
work in the Milne co-ordinate system where, 
\begin{align}
\tau &= \sqrt{t^2-z^2}, \label{MC1}\\
\eta_s &= \tanh^{-1}(z/t), \label{MC2}\\
r &= \sqrt{x^2+y^2}, \label{MC3}\\
\varphi &= {\rm atan2}(y,x). \label{MC4}
\end{align}
The metric tensor for this co-ordinate system is $g_{\mu\nu}={\rm 
diag}(1,\,-\tau^2,\,-1,\,-r^2)$. Boost invariance and rotational 
invariance implies $u^\varphi=u^{\eta_s}=0$, whereas linearly rising 
transverse velocity flow profile leads to $u^r\sim r$. The 
blast-wave model further assumes that the particle freeze-out 
happens at a proper time $\tau_f$ having a constant temperature $T_f$
and uniform matter distribution, in the transverse plane. In 
summary, the hydrodynamic fields are parametrized as \cite 
{Teaney:2003kp}
\begin{align}
T &= T_f\, \Theta(R-r), \label{BW1}\\
u^r &= u_0\,\frac{r}{R}\, \Theta(R-r) \label{BW2}\\
u^\varphi &= u^{\eta_s}= 0, \label{BW3}\\
u^\tau &= \sqrt{1 + (u^r)^2}, \label{BW4}
\end{align}
where $R$ is the transverse radius of the fireball at freeze-out. 
The expression for $u^\tau$ is obtained by requiring that the fluid 
four-velocity satisfy the condition $u^\mu u_\mu=1$.

The hadron spectra can be obtained using the Cooper-Frye 
prescription for particle production \cite{Cooper:1974mv}
\begin{equation}\label{CF}
\frac{dN}{d^2p_Tdy} = \frac{1}{(2\pi)^3}\int p_\mu d\Sigma^\mu f(x,p),
\end{equation}
where $d\Sigma_\mu$ is the oriented freeze-out hyper-surface and 
$f(x,p)$ is the phase-space distribution function of the particles 
at freeze-out. The distribution function can be written in terms of 
the equilibrium and non-equilibrium parts, $f=f_0+\delta f$. The 
equilibrium distribution function is given by
\begin{equation}\label{IDF}
f_0 = \frac{1}{\exp(u_\mu p^\mu/T) + a}~,
\end{equation}
where $a=+1$ for baryons and $a=-1$ mesons. 

For small deviations from equilibrium, i.e., $\delta f\ll f_0$, we 
use the Grad's 14-moment approximation for the non-equilibrium part 
\cite{Grad, Romatschke:2009im}
\begin{equation}\label{Grad}
\delta f = \frac{f_0\tilde f_0}{2(\epsilon+P)T^2}\, p^\alpha p^\beta \pi_{\alpha\beta},
\end{equation}
where $\tilde f_0 = 1 - a f_0$ and $\pi_{\alpha\beta}$ is the shear 
stress tensor. Approximating the shear stress tensor with its 
first-order relativistic Navier-Stokes expression, 
$\pi_{\alpha\beta}= 2\eta\nabla_{\langle\alpha} u_{\beta\rangle}$, 
the expression for the 14-moment approximation reduces to 
\begin{equation}\label{GradT}
\delta f_1 = \frac{f_0\tilde f_0}{T^3}\left(\frac{\eta}{s}\right)p^\alpha p^\beta \nabla_{\langle\alpha} u_{\beta\rangle}.
\end{equation}
Here $\eta$ is the coefficient of shear viscosity, $s=(\epsilon+P)/T$
is the entropy density and the angular brackets denote traceless 
symmetric projection orthogonal to the fluid four-velocity \cite 
{Bhalerao:2013pza}. The form of $p^\alpha p^\beta 
\nabla_{\langle\alpha}u_{\beta\rangle}$ in the case of blast-wave 
model is calculated in Appendix~\ref{Appendix A}.

The anisotropic flow is defined as
\begin{equation}\label{vn}
v_n(p_T) \equiv 
\dfrac{\displaystyle{\int_{-\pi}^{\pi}}d\phi\,\cos[n(\phi-\Psi_n)]\,\dfrac{dN}{dy\,p_t\,dp_T\,d\phi}}
{\displaystyle{\int_{-\pi}^{\pi}}d\phi\,\dfrac{dN}{dy\,p_t\,dp_T\,d\phi}},
\end{equation}
where $\Psi_n$ is the $n$-th harmonic event-plane angle. In the 
present case, we do not consider event-by-event fluctuations and 
therefore $\Psi_n=0$. Up to first order in viscosity \cite 
{Teaney:2003kp},
\begin{align}\label{vnFO}
v_n(p_T) =&\, v_n^{(0)}(p_T)\left(1 - \dfrac{\displaystyle{\int d\phi\,\dfrac{dN^{(1)}}{dy\,p_t\,dp_T\,d\phi}}}
{\displaystyle{\int d\phi\,\dfrac{dN^{(0)}}{dy\,p_t\,dp_T\,d\phi}}}\right) \nonumber\\
&+ \dfrac{\displaystyle{\int d\phi\,\cos[n(\phi-\Psi_n)]\,\dfrac{dN^{(1)}}{dy\,p_t\,dp_T\,d\phi}}}
{\displaystyle{\int d\phi\,\dfrac{dN^{(0)}}{dy\,p_t\,dp_T\,d\phi}}},
\end{align}
where the superscript `$(0)$' denotes quantities calculated using 
the ideal distribution function, Eq.~(\ref{IDF}), and `$(1)$' 
denotes those obtained using the first-order viscous correction, 
Eq.~(\ref{GradT}).


\section{Viscous blast-wave model}

The definition of the participant anisotropies, $\varepsilon_n$, via 
the Fourier expansion for a single-particle distribution is
\begin{equation}\label{RFP}
f(\varphi) = \frac{1}{2\pi}\left[1 + 2\sum_{n=1}^\infty\,\varepsilon_n\,\cos[n(\varphi-\psi_n)]\right],
\end{equation}
where $\psi_n$ are the angles between the $x$ axis and the major 
axis of the participant distribution. The geometrical anisotropies 
in the initial particle distribution, $\varepsilon_n$, eventually 
converts to anisotropies in the radial fluid velocity,
\begin{equation}\label{RFP}
u^r = u_0\,\frac{r}{R}\left[1 + 2\sum_{n=1}^\infty\,u_n\,\cos[n(\varphi-\psi_n)]\right].
\end{equation}
In the following, we determine the conversion efficiency of the 
initial eccentricity to anisotropy in the radial fluid velocity, 
$u_n/\varepsilon_n$.

Using the well known dispersion relation for sound in a viscous 
medium \cite{Romatschke:2009im},
\begin{equation}\label{cs_disp}
\omega = c_s k + ik^2 \frac{1}{T}\left(\frac{2}{3}\frac{\eta}{s}\right),
\end{equation}
the authors of Ref.~\cite{Staig:2010pn} introduced a viscosity-based 
survival scale which all structures formed by point like 
perturbations should attain at freeze-out. In the above equation, 
$\eta$ is the coefficients of shear viscosity and $c_s$ is the speed 
of sound in the medium. In the present work, we ignore the 
contribution due to bulk viscosity. Using a plane-wave Fourier 
ansatz, $\exp(i\omega t -ikx)$, we observe that the amplitudes of 
the stress tensor harmonics with momentum $k$ are attenuated by a 
factor
\begin{equation}\label{VBSS}
\delta T^{\mu\nu}(t,k) = \exp\left[-\left(\frac{2}{3}\frac{\eta}{s}\right)
\frac{k^2t}{T}\right]\delta T^{\mu\nu}(0,k),
\end{equation}
where we have suppressed the oscillatory pre-factor. We note that 
the presence of momentum squared in the exponent leads to enhanced 
effect of viscosity for the higher harmonics. We expect the same 
qualitative behaviour for the radial flow velocity as will be 
explained in the following.

\begin{figure}[t]
\begin{center}
\includegraphics[width=\linewidth]{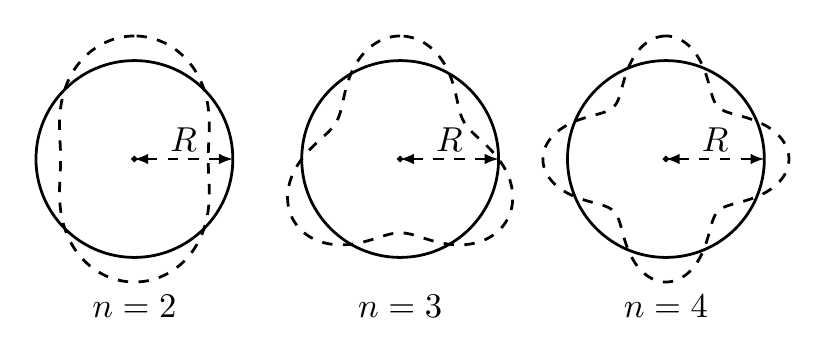}
\end{center}
\vspace{-0.6cm}
\caption{The harmonics form standing waves on the fireball 
circumference having radius $R$ at freezeout.}
\label{sine_circle}
\end{figure}

First and foremost, we notice that each harmonics is essentially a 
damped oscillator with wave-vector $k$. Moreover, throughout the 
evolution the harmonics form standing waves on the fireball 
circumference, as shown in Fig.~\ref{sine_circle}, whose amplitude 
is progressively damped due the viscous effects. Therefore, the 
fireball circumference is an integer multiple of the wavelength with 
wave-vector $k$, i.e.,
\begin{equation}\label{SWH}
2\pi R = n\frac{2\pi}{k},
\end{equation}
where $R$ is the transverse radius of the fireball at freeze-out. 
Hence, at the freeze-out time $t_f$, the wave amplitude reaction is 
given by
\begin{equation}\label{VBSS}
\frac{\delta T^{\mu\nu}|_{t=t_f}}{\delta T^{\mu\nu}|_{t=0}} = 
\exp\left[-n^2\left(\frac{2}{3}\frac{\eta}{s}\right)\frac{t_f}{R^2T_f}\right],
\end{equation}
where $T_f$ is the freeze-out temperature. 

In absence of viscosity, the initial geometrical perturbations in 
the fluid will result in the development of radial flow velocity and 
the conversion efficiency will remain the same for all harmonics. In the 
case of a viscous medium, however, the conversion efficiency of the initial 
geometrical perturbation to radial fluid velocity must be 
proportional to the wave amplitude reaction,
\begin{equation}\label{VBSS}
\frac{u_n}{\varepsilon_n} = \alpha_0
\exp\left[-n^2\left(\frac{2}{3}\frac{\eta}{s}\right)\frac{t_f}{R^2T_f}\right],
\end{equation}
where $\alpha_0$ is the constant of proportionality. The sudden 
stopping of the damped oscillator at the freeze-out time may lead to 
certain phases due to the oscillatory pre-factor. These phases can, 
in general, lead to secondary peaks in the power spectrum of higher 
harmonics. However, as no secondary peaks has been observed in the 
spectrum of relativistic heavy-ion collisions, we will continue to 
ignore these phase factors.
  
We emphasize that the acoustic damping should be applied to the 
hydrodynamic variables, such as the moments of the flow velocity, 
$u_n$, rather then to the final state observables such as $v_n$, as 
done in Ref.~\cite{Shuryak:2013ke}. In Ref.~\cite{Shuryak:2013ke}, 
Shuryak and Zahed (S-Z) proposed that the ratio of the initial 
eccentricity $\varepsilon_n$ to the final $p_T$-integrated $v_n$ 
should be proportional to the wave amplitude reaction, i.e., the 
r.h.s. of Eq.~(\ref{VBSS}) should be equal to $v_n/\varepsilon_n$. 
However, $\varepsilon_n$ is the eccentricity in the configuration 
space whereas $v_n$ is the momentum anisotropy of the particles 
after freeze-out. The momentum space asymmetries are not 
hydrodynamic variables and are only affected indirectly via damping. 
On the other hand, the acoustic damping should be applicable to 
hydrodynamic variables and it should only capture the viscous 
effects of the hydrodynamic evolution. This assumption also misses 
the additional effect of viscosity at freeze-out using the 
Cooper-Frye formula. Moreover, it does not provide the opportunity 
to study the $p_T$ dependence of anisotropic flow and hence to 
estimate the viscosity of the expanding medium.


\section{Initial conditions}

In this section, we set-up the initial conditions of the collisions 
in order to reduce the number of free parameters in the blast-wave 
model. To this end, we evaluate the parameters corresponding to the 
initial geometry of the collisions. We also approximate the 
subsequent transverse expansion of the fireball by using the radial 
velocity parametrization in the blast-wave model.

\begin{figure}[t]
\begin{center}
\includegraphics[width=\linewidth]{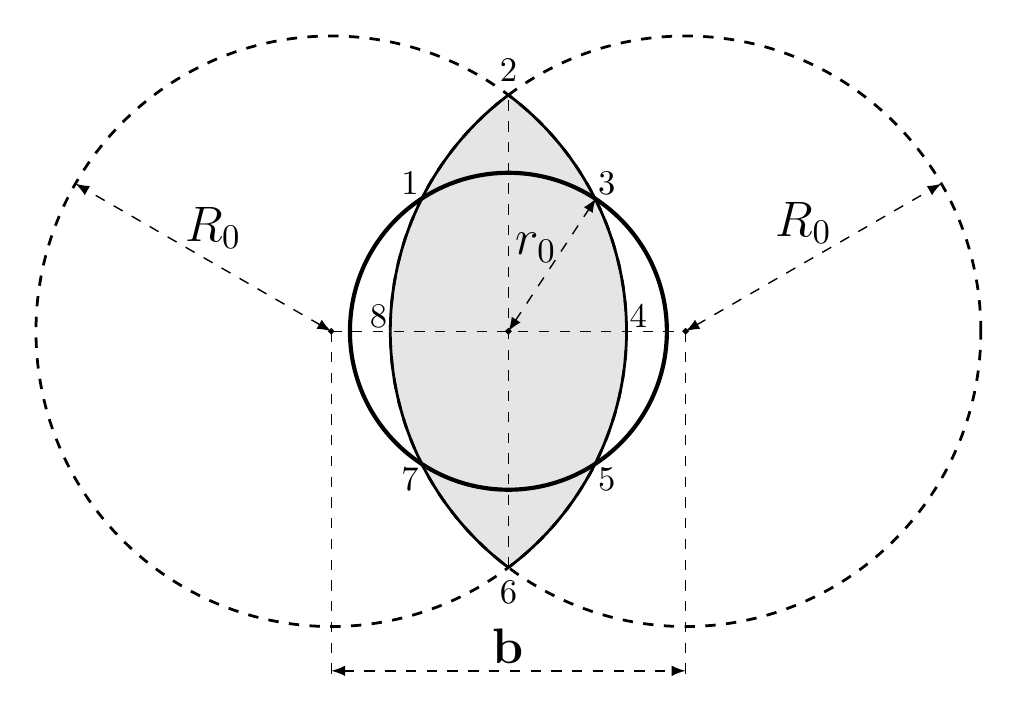}
\end{center}
\vspace{-0.4cm}
\caption{Initial transverse geometry of the collision of two 
identical nuclei with radius $R_0$ and impact parameter $b$. The 
shaded region represents the the overlap zone of the colliding 
nuclei. The circle with radius $r_0$ is drawn such that it equally 
divides the boundary of the overlap zone in four parts, i.e., 
$\wideparen{123}=\wideparen{345}=\wideparen{567}=\wideparen{781}$.}
\label{ini_geo}
\end{figure}

We consider the collision of two identical nuclei with mass number 
$A$. The radius of each nucleus is given by $R_0=1.25A^{1/3}$ fm and 
the impact parameter is $b$, as shown in Fig.~\ref{ini_geo}. The 
shaded region in Fig.~\ref{ini_geo} is the overlap zone of the 
colliding nuclei. We draw a circle of radius $r_0$, with its centre 
coinciding with that of the overlap zone, in such a way that the 
boundary of the overlap zone is equally divided in four parts, i.e., 
$\wideparen{123}=\wideparen{345}=\wideparen{567}=\wideparen{781}$; 
see Fig.~\ref{ini_geo}. This approximates $\varepsilon_2$ as the 
second harmonics of initial geometrical fluctuations, analogous to 
the $n=2$ case as shown in Fig.~\ref{sine_circle}. The radius $r_0$ 
is therefore the initial transverse radius of the expanding fireball 
and is given by
\begin{equation}\label{r0}
r_0 = \frac{1}{2}\left( b^2 - 2\,bR_0\sqrt{2+\frac{b}{R_0}} + 4R_0^2\right)^{1/2},
\end{equation}
which reduces to $r_0=R_0$ for head-on collisions ($b=0$). Since 
$\varepsilon_2$ is the most prominent eccentricity for non-central 
collisions, all other geometrical eccentricities are treated as 
boundary perturbations to this circle.

The subsequent transverse expansion of the fireball is obtained by 
employing the radial velocity parametrization of the blast-wave 
model. Using the perturbation-free expression for the transverse 
velocity, Eq.~(\ref{BW2}), we get
\begin{equation}\label{FRF}
u^r \equiv \frac{dr}{d\tau} = u_0\frac{r}{R} ~~\Rightarrow~
\int_{r_0}^R\frac{dr}{r} = \int_0^{\tau_f}\frac{u_0}{R}\,d\tau.
\end{equation} 
After performing the straightforward integration, we obtain a 
transcendental equation for the freeze-out radius, $R$,
\begin{equation}\label{FOR}
R = r_0 \exp\left(\frac{u_0\,\tau_f}{R}\right),
\end{equation}
which can be solved for $R$ given the isotropic expansion velocity 
$u_0$ and the freeze-out time $\tau_0$. In the following, using 
Bjorken's scaling solution for one-dimensional boost-invariant 
expansion, we obtain an expression to determine the freeze-out times 
for non-central collisions once it is fixed for the central one.

\begin{figure}[t]
\begin{center}
\includegraphics[width=\linewidth]{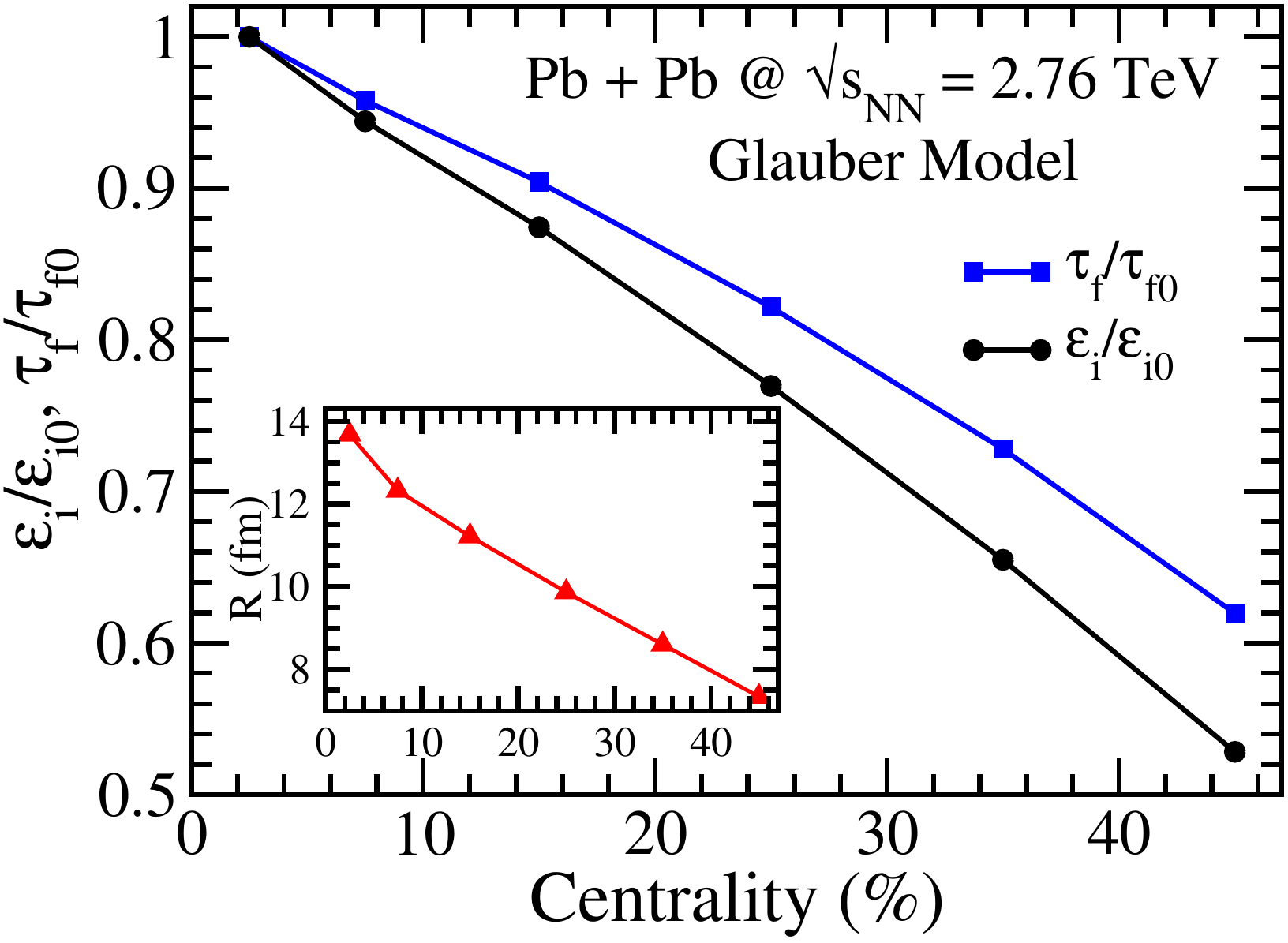}
\end{center}
\vspace{-0.4cm}
\caption{(Color online) Centrality dependence of the initial energy 
density $\epsilon_i$ and freeze-out time $\tau_f$ for Pb+Pb 
collisions at $\sqrt{s_{NN}}=2.76$ TeV scaled by the corresponding 
values in $0-5$\% central collisions. The inset shows centrality 
dependence of the freeze-out radius of the fireball $R$, obtained 
using Eq.~(\ref{FOR}).}
\label{eiei0}
\end{figure}

For the ideal hydrodynamic evolution of relativistic fluid, in the 
one-dimensional boost-invariant scenario, the evolution of the 
energy density follows $\epsilon\propto\tau^{-4/3}$. Assuming the 
initial thermalization time and the final freeze-out energy density 
(i.e., the freeze-out temperature) to be same for all collisions, we 
get
\begin{equation}\label{FOT}
\tau_f = \tau_{f0}\left(\frac{\epsilon_i}{\epsilon_{i0}}\right)^{3/4}.
\end{equation}
Here $\tau_{f0}$ is the freeze-out time for most central collisions, 
which has to be fixed by fitting the corresponding transverse 
momentum spectra. The freeze-out times for other centralities can 
then be obtained using the above equation and therefore they are not 
free parameters. The ratio $\epsilon_i/ \epsilon_{i0}$ is the 
initial central energy density scaled by its corresponding value in 
most central collisions. 

Figure~\ref{eiei0} shows $\epsilon_i/\epsilon_{i0}$ for various 
centralities obtained using the Glauber model calculations \cite
{Kolb:2001qz, Roy:2011xt} in the case of Pb--Pb collisions at 
$\sqrt{s_{NN}}=2.76$ TeV at the LHC. We observe that 
$\epsilon_i/\epsilon_{i0}$ (and hence the initial temperature) 
decreases for non-central collisions compared to central ones. 
Therefore according to Eq.~(\ref{FOT}), freeze-out happens earlier 
in peripheral collisions which is also reflected in Fig.~\ref{eiei0} 
for $\tau_f/\tau_{f0}$. The inset of Fig.~\ref{eiei0} shows the 
centrality dependence of the freeze-out radius of the fireball $R$, 
obtained using Eq.~(\ref{FOR}). We find a rather large transverse 
radius of the fireball at freeze-out. Finally, the parameters that 
we need to fix within the viscous-blast wave model to fit the 
spectra are the freeze-out temperature $T_f$, the freeze-out time 
for central collision $\tau_{f0}$ and the unperturbed maximum radial 
flow velocity $u_0$. An interplay of the coefficient of 
proportionality $\alpha_0$ in Eq.~(\ref{VBSS}) and $\eta/s$ will be 
crucial to reproduce the flow harmonics.


\section{Results and discussions}

In this section, we show our results for Pb+Pb collisions at 
$\sqrt{s_{NN}}=2.76$ TeV and compare it with experimental data 
measured at LHC by the ALICE and ATLAS collaborations.

\begin{figure}[t]
\begin{center}
\includegraphics[width=\linewidth]{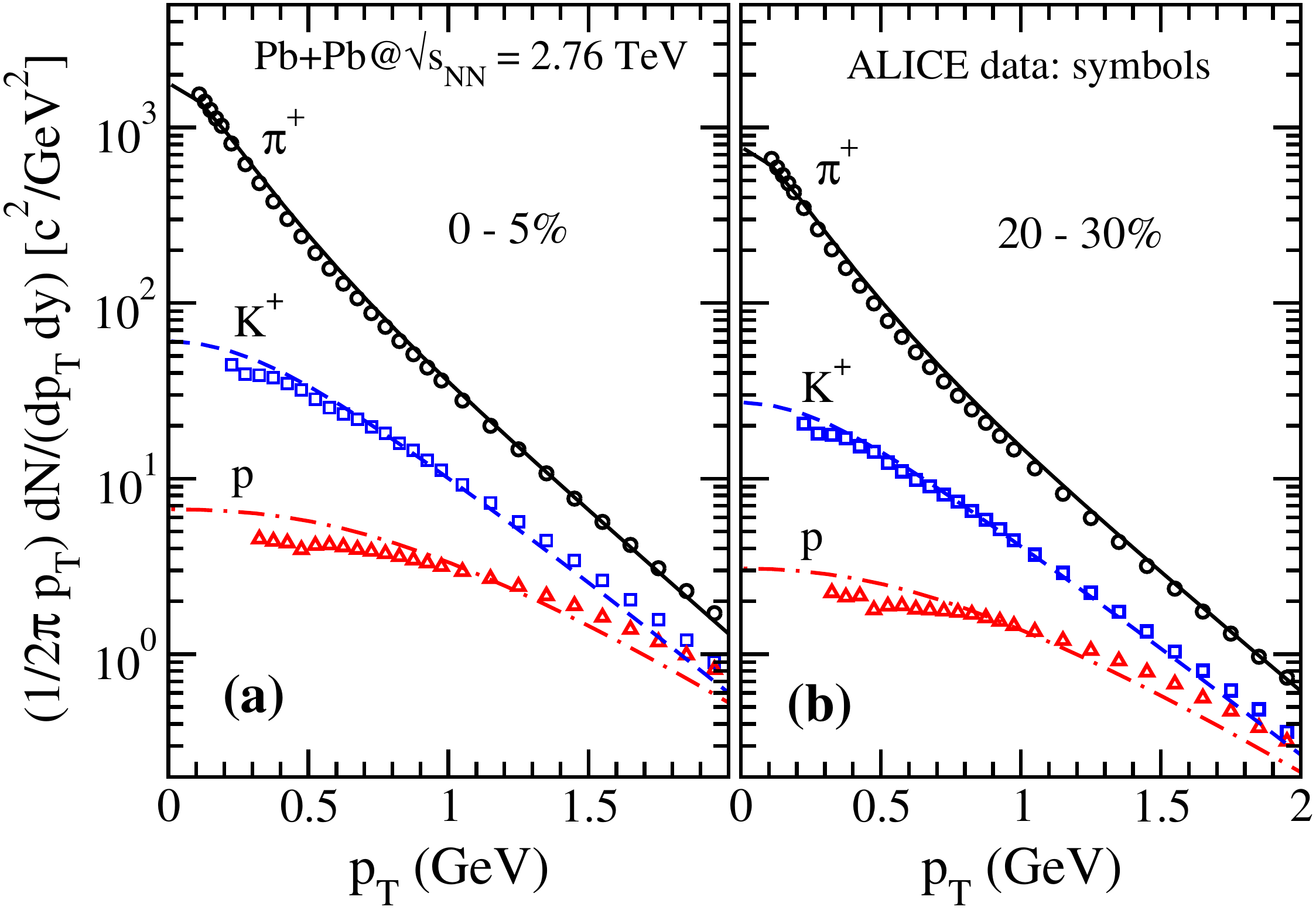}
\end{center}
\vspace{-0.4cm}
\caption{(Color online) Transverse momentum distribution of particle 
multiplicities in Pb+Pb collisions at $\sqrt{s_{NN}}=2.76$ TeV. We 
show results for $\pi^+$, $K^+$, and $p$ in two centrality ranges, 
(a): $0-5$\% and (b): $20-30$\%. The symbols represent ALICE data 
\cite{Abelev:2013vea} at midrapidity and the lines correspond to 
viscous blast-wave calculations.}
\label{hadpt}
\end{figure}

Figure \ref{hadpt} shows the transverse momentum distribution of 
pions, kaons, and protons spectra for $0-5$\% and $20-30$\% central 
Pb+Pb collisions at $\sqrt{s_{NN}}=2.76$ TeV measured at LHC by the 
ALICE collaboration (symbols) at midrapidity \cite{Abelev:2013vea} 
and calculated using the viscous blast-wave model (lines). The 
results are obtained using root mean square values of eccentricities 
$\varepsilon_2$ and $\varepsilon_3$ in a Monte-Carlo Glauber model, 
with a shear viscosity to entropy density ratio $\eta/s =0.24$. We 
observe that the spectra for $\pi^+$ and $K^+$ from the viscous 
blast-wave model are in good overall agreement with the experimental 
data for a freeze-out temperature of $120$ MeV. On the other hand, 
within the viscous blast-wave model, the proton yield for a 
freeze-out temperature of $120$ MeV is severely underestimated. To 
obtain an overall fair agreement with the experimental data, the 
freeze-out temperature for protons is considered to be $135$ MeV. 
The freeze-out time for $0-5$\% most central collision was found to 
be $8$ fm. For other centralities, the freeze-out time was obtained 
by using Eq.~(\ref{FOT}) and results shown in Fig.~\ref{eiei0}.

\begin{figure}[t]
\begin{center}
\includegraphics[width=\linewidth]{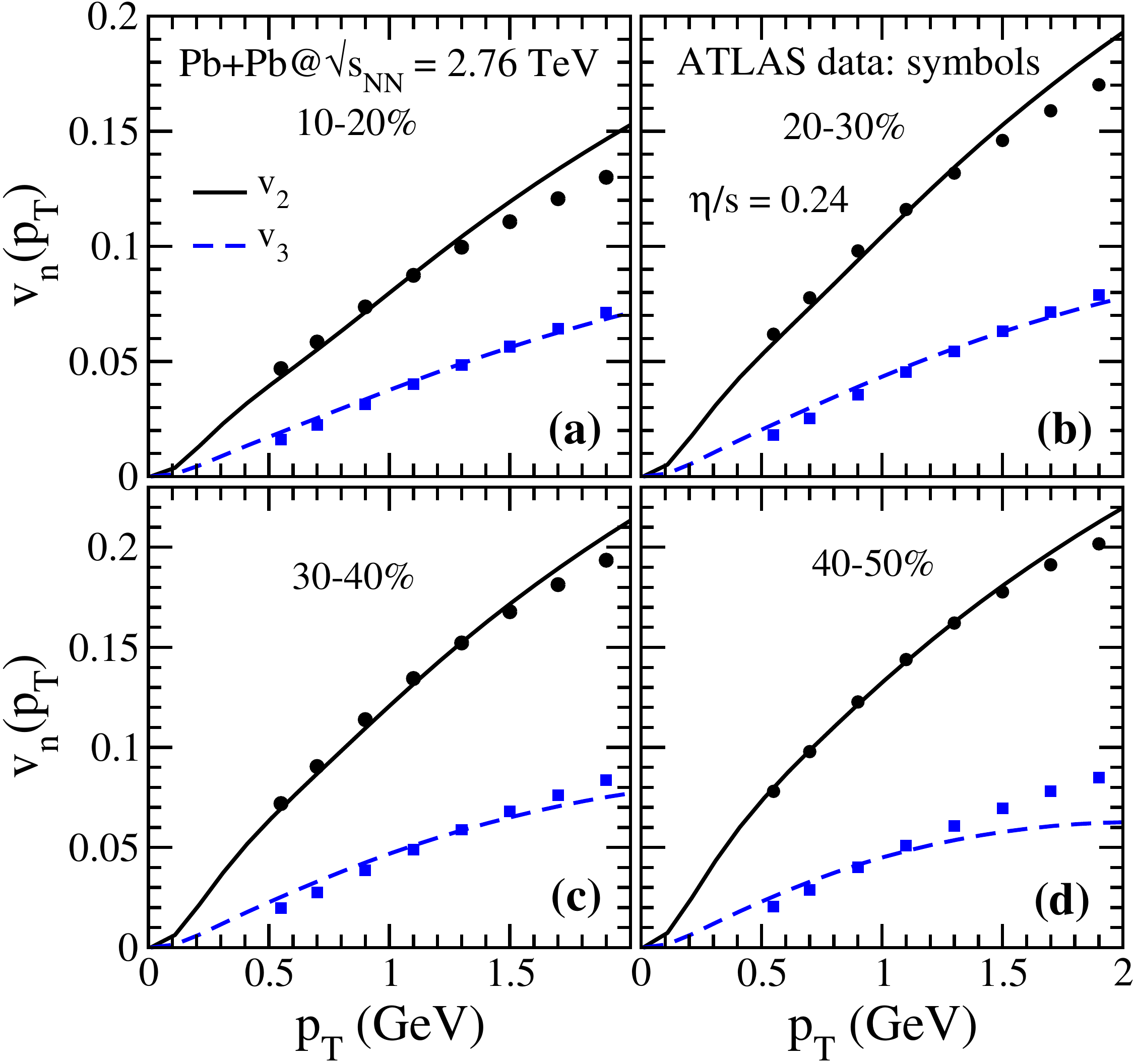}
\end{center}
\vspace{-0.4cm}
\caption{(Color online) Transverse momentum dependence of the 
anisotropic flow coefficients $v_n(p_T)$ of charged hadrons, for 
$n=2$ and $3$, calculated at various centralities in Pb+Pb 
collisions at $\sqrt{s_{NN}}=2.76$ TeV in the viscous blast-wave 
model (lines) with $\eta/s=0.24$ as compared to the ATLAS data \cite 
{ATLAS:2012at} (symbols).}
\label{vnpT}
\end{figure}

Figure~\ref{vnpT} shows our results for the $v_n(p_T)$, in 
comparison with the ATLAS data \cite{ATLAS:2012at} for various 
centralities. We find overall fair agreement with the data for 
elliptic ($v_2$) and triangular ($v_3$) flow, at all centralities. 
This is achieved by choosing a single fixed value $\eta/s = 0.24$ to 
obtain the required suppression (relative to ideal blast-wave 
results) of $v_2$ and $v_3$, for all centralities. Apart from 
$\eta/s$, flow also depends on the constant of proportionality 
$\alpha_0$ appearing in Eq.~(\ref{VBSS}). It controls the conversion 
efficiency of the initial eccentricity to final fluid velocity. We 
consider the initial eccentricity $\varepsilon_n$ to be the rms 
values of the eccentricities obtained in the MC-Glauber model, as 
given in Ref.~\cite{Retinskaya:2013gca}. A large value of $\alpha_0$ 
means larger conversion of eccentricity leading to increased flow 
velocity and hence higher $v_2$ and $v_3$. On the other hand, as is 
well known, an increase in $\eta/s$ leads to suppression of $v_2$ 
and $v_3$. Large (small) value of $\alpha_0$ can be compensated by 
choosing a higher (lower) value of $\eta/s$ up to a certain extent. 
However, beyond a certain range of $\eta/s$, the relative behaviour 
of $v_2$ and $v_3$ is destroyed. 

In order to match the elliptic and triangular flow data, we find the 
most suitable parameter values for $\alpha_0=0.4$ and $\eta/s=0.24$. 
Moreover, we find that $v_n$ is insensitive to $a$ for transverse 
velocity of the form $v^r\sim (r/R)^a$. This may be attributed to 
the fact that the exponent $a$ controls the rate of isotropic 
transverse expansion. The anisotropic flow originates from the 
initial eccentricity which translates into final flow. Therefore 
$v_n$ is sensitive to $u_n$ which depends on $\alpha_0$ and $\eta/s$, 
as is apparent from Eq.~(\ref{VBSS}). On the other hand, it should 
be noted that the slope of the particle spectra is sensitive to the 
exponent $a$ and could be tuned to get a better fit with the 
experimental data. However, in the present work, we are interested 
in the anisotropic flow and therefore we set $a=1$ and do not fit it 
to match the particle spectra.

\begin{figure}[t]
\begin{center}
\includegraphics[width=\linewidth]{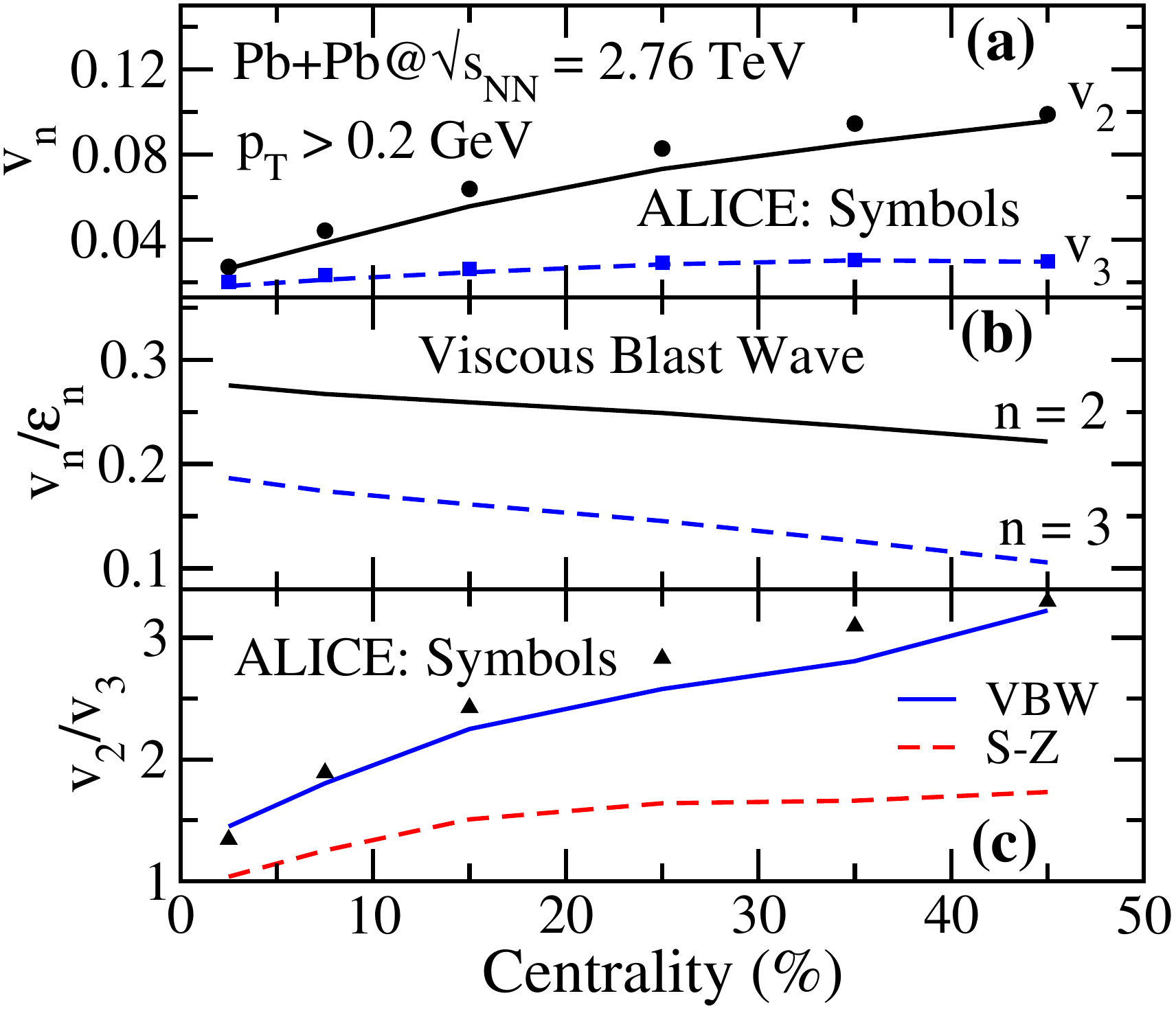}
\end{center}
\vspace{-0.4cm}
\caption{(Color online) (a): Centrality dependence of the $p_T$ 
integrated anisotropic flow coefficients $v_n$ of charged hadrons in 
Pb+Pb collisions at $\sqrt{s_{NN}}=2.76$ TeV calculated in the 
viscous blast-wave model (lines) with $\eta/s =0.24$, as compared to 
ALICE data \cite{ALICE:2011ab} (symbols). (b): Centrality dependence 
of the ratio $v_n/\varepsilon_n$ in the viscous blast-wave model 
where $\varepsilon_n$ is the rms values of the eccentricities 
obtained in the MC-Glauber model \cite{Retinskaya:2013gca}. In 
panels (a) and (b), we show results for $n=2$ and $3$. (c): 
Centrality dependence of the ratio $v_2/v_3$ for ALICE data 
(symbols), using the viscous blast-wave model (solid line) and from 
the Shuryak-Zahed estimate (dashed line).} 
\label{vn_en}
\end{figure}

In Fig.~\ref{vn_en}(a), $p_T$-integrated values of $v_2$ and $v_3$ 
obtained from the viscous blast-wave model are compared with ALICE 
data \cite{ALICE:2011ab}, as a function of centrality. We observe 
that using the same constant $\eta/s=0.24$ for all centralities, the 
model shows a good agreement with the data. Since $v_2$ is driven 
mostly by the initial spatial anisotropy, it exhibits a strong 
centrality dependence compared to $v_3$. Figure~\ref{vn_en}(b) shows 
the conversion efficiency of the initial spatial anisotropy into the 
final momentum anisotropy. A linear relation between $v_n$ and 
$\varepsilon_n$, as observed in some hydrodynamic calculations \cite 
{Bhalerao:2015iya}, is not obtained within the viscous blast-wave 
model presented here. On the other hand, in view of the non-linear 
nature of the hydrodynamic equations a linear relation between $v_n$ 
and $\varepsilon_n$ is not obvious. Indeed other calculations \cite 
{Niemi:2012aj} as well as a recent analysis of the LHC data \cite 
{Yan:2014nsa} also result in a similar centrality dependence as 
obtained in our analysis In Fig.~\ref{vn_en}(c), we show the ratio 
$v_2/v_3$ as a function of centrality for the ALICE data (symbols), 
the present viscous blast-wave model (blue solid line) and the 
estimate due to Shuryak and Zahed (red dashed line). We see that the 
viscous blast-wave model provides a better agreement with ALICE data 
compared to the S-Z estimate. However, we should keep in mind that 
the freeze-out parameters used in the S-Z estimate is the same as 
that of the viscous blast-wave fit values.


\section{Conclusions and outlook}

In this paper we have generalized the blast-wave model to include 
viscous effects by employing a viscosity-based survival scale for 
geometrical anisotropies formed in the early stages of relativistic 
heavy-ion collisions. This viscous damping is introduced in the 
parametrization of the radial flow velocity. The viscous blast wave 
model presented here involved five parameters, including $\eta/s$, 
which has to be fitted for only one centrality. This model therefore 
incorporates the important features of viscous hydrodynamic 
evolution but does not require to do the actual evolution. We have 
used this viscous blast-wave model to obtain the transverse momentum 
distribution of particle yields and anisotropic flow harmonics for 
LHC. The blast-wave model parameters were fixed by fitting the 
transverse momentum spectra of identified particles. We demonstrated 
that a fairly good agreement was achieved for transverse momentum 
distribution of elliptic and triangular flow for various 
centralities as well as centrality distribution for the integrated 
flow. Within the present model, we estimated the shear viscosity to 
entropy density ratio $\eta/s\simeq 0.24$ at the LHC.

One of the drawbacks of the present model is that we have employed 
root mean squared eccentricity over number of events, which is 
analogous to ``single shot" hydrodynamic evolution. On the plus 
side, the present model could also be implemented on an 
event-by-event basis. Another difficulty that we encountered was 
that in order to fit the proton spectra, we had to consider a 
different freeze-out temperature ($135$ MeV) compared to that for 
pions and Kaons ($120$ MeV). This problem could be addressed by 
parametrizing the transverse velocity in the form $v^r\sim (r/R)^a$ 
and fitting $a$ for different particle species, separately. We leave 
these problems for a future work.


\begin{acknowledgments}

A.J. acknowledges useful discussions with Subrata Pal, Krzysztof 
Redlich, Edward Shuryak and Derek Teaney. A.J. thanks Victor Roy for 
helpful comments and providing data from Glauber model. A.J. was 
supported by the Frankfurt Institute for Advanced Studies (FIAS), 
Germany.

\end{acknowledgments}


\appendix

\section{Viscous stress tensor}
\label{Appendix A}

In this Appendix, we calculate the viscous tensor 
$\nabla_{\langle\alpha}u_{\beta\rangle}$ and therefore obtain the 
viscous corrections to the distribution function at freeze-out. We 
work in Milne co-ordinate system, Eqs.~(\ref{MC1})-(\ref{MC4}) with 
the metric tensor $g_{\mu\nu}={\rm diag}(1,\,-\tau^2,\,-1,\,-r^2)$. 
Therefore, the inverse metric tensor is $g^{\mu\nu}={\rm 
diag}(1,\,-1/\tau^2,\,-1,\, -1/r^2)$, its determinant $g$ is 
$\sqrt{-g}=\tau r$ and the non-vanishing Christoffel symbols are 
$\Gamma^\tau_{\eta_s\eta_s}=\tau$, $\Gamma^{\eta_s}_{\tau\eta_s} 
=1/\tau$, $\Gamma^r_{\varphi\varphi}=-r$, and 
$\Gamma^\varphi_{r\varphi}=1/r$. Using the parametrization of the 
fluid velocity given in Eqs.~(\ref {BW2})-(\ref{BW4}), we get
\begin{align}
\Delta^{r\varphi} &= 0, \quad
\Delta^{\varphi\varphi} = -\frac{1}{r^2}, \quad
\Delta^{rr} = - 1 - (u^r)^2, \label{identity1}\\
\partial_r u^r &= \frac{u^r}{r}, \quad
\partial_\varphi u^r = -\frac{2u_0\,r}{R}\sum_{n=1}^{\infty}n\,u_n\sin[n(\varphi-\psi_n)].\label{identity2}
\end{align}
where $\Delta^{\mu\nu}\equiv g^{\mu\nu}-u^\mu u^\nu$ is the 
projection operator orthogonal to the fluid four-velocity. 

To fix the time derivatives of the fluid velocity, we assume that if 
the particles are freezing-out, they are free streaming, which means 
that $Du^\mu=0$. Here $D\equiv u^\mu d_\mu$ is the co-moving 
derivative and $d_\mu$ is the covariant derivative. With this 
prescription, we have
\begin{align}\label{TD}
\partial_\tau u^\varphi &= 0, \nonumber\\
\partial_\tau u^r &= -v\,\partial_r u^r = -\frac{(u^r)^2}{ru^\tau}, \nonumber\\
\partial_\tau u^\tau &= v\,\partial_\tau u^r = - \frac{(u^{r})^3}{r(u^\tau)^2}
\end{align}
where $v=u^r/u^\tau$ is the radial velocity. The expansion scalar is 
given by
\begin{align}\label{ES}
\frac{1}{\sqrt{-g}} \partial_{\mu} (\sqrt{-g} u^{\mu}) &= 
\frac{u^{\tau}}{\tau} + \frac{u^{r}}{r} + \partial_\varphi u^\varphi + \partial_r u^r + \partial_\tau u^\tau, \nonumber\\
&= \frac{u^{\tau}}{\tau} + 2\frac{u^{r}}{r} - \frac{(u^{r})^3}{r(u^\tau)^2}.
\end{align}
Assuming boost invariance, the spatial components of the viscous 
tensor are given by
\begin{align}
r \nabla^{\langle r} u^{\varphi\rangle} =&\, -\frac{r}{2} \partial_r u^{\varphi} - \frac{1}{2r}\partial_{\varphi} {u^{r}} 
- \frac{r}{2}u^r Du^\varphi - \frac{r}{2}u^\varphi Du^r \nonumber\\
&- \frac{1}{3}r\Delta^{r\varphi}\frac{1}{\sqrt{-g}} \partial_{\mu}(\sqrt{-g} u^\mu ) \nonumber\\
=&~ \frac{u_0}{R}\sum_{n=1}^{\infty}n\,u_n\sin[n(\varphi-\psi_n)]\,, \label{SRP}
\end{align}
\begin{align}
r^2\nabla^{\langle\varphi}u^{\varphi\rangle} =&\, -\partial_\varphi u^{\varphi} - \frac{u^{r}}{r} - r^2 u^{\varphi} Du^{\varphi} \nonumber\\
&-\frac{1}{3} r^2 \Delta^{\varphi\varphi} \frac{1}{\sqrt{-g}} \partial_{\mu} (\sqrt{-g} u^{\mu} ) \nonumber\\
=&~ \frac{1}{3}\left[\frac{u^{\tau}}{\tau} - \frac{u^{r}}{r} - \frac{(u^{r})^3}{r(u^\tau)^2}\right]\,, \label{SPP}
\end{align}
\begin{align}
\nabla^{\langle r}u^{r\rangle} =&\, - \partial_r u^{r} - u^{r} Du^{r} 
- \frac{1}{3} \Delta^{rr} \frac{1}{\sqrt{-g}} \partial_{\mu} (\sqrt{-g} u^{\mu} ) \nonumber\\
=&~ \frac{1}{3}\left[\frac{(u^{\tau})^3}{\tau} - \frac{u^{r}}{r} + \frac{(u^{r})^3}{r}\right]\,, \label{SRR}
\end{align}
where we have used the fact that $(u^\tau)^2 = 1 + (u^r)^2$.
\begin{align}
\tau^2\nabla^{\langle\eta_s}u^{\eta_s\rangle} =&\, - \frac{u^{\tau}}{\tau}  
+ \frac{1}{3} \frac{1}{\sqrt{-g}} \partial_{\mu} (\sqrt{-g} u^{\mu} ) \nonumber\\
=&~ \frac{1}{3}\left[2\frac{u^{r}}{r} - 2\frac{u^{\tau}}{\tau} - \frac{(u^{r})^3}{r(u^\tau)^2}\right]\,, \label{SEE}
\end{align}
\begin{align}
\nabla^{\langle r}u^{\eta_s\rangle} =&~ \nabla^{\langle\varphi}u^{\eta_s\rangle} = 0  \label{SRESPE}\; .
\end{align} 

To obtain the temporal components of the viscous stress energy 
tensor, we use the Landau frame condition, 
$\nabla^{\langle\alpha}u^{\beta\rangle}u_{\beta}=0$.
\begin{align}
\nabla^{\langle\tau}u^{\tau\rangle}u_\tau + \nabla^{\langle\tau}u^{r\rangle}u_r &= 0 
~~\Rightarrow~~ \nabla^{\langle\tau}u^{\tau\rangle} = v\nabla^{\langle\tau}u^{r\rangle}, \label{STT}\\
\nabla^{\langle\eta_s}u^{\tau\rangle}u_\tau + \nabla^{\langle\eta_s}u^{r\rangle}u_r &= 0 
~~\Rightarrow~~ \nabla^{\langle\tau}u^{\eta_s\rangle} = 0, \label{STE}\\
\nabla^{\langle r}u^{\tau\rangle}u_\tau + \nabla^{\langle r}u^{r\rangle}u_r &= 0 
~~\Rightarrow~~ \nabla^{\langle\tau}u^{r\rangle} = v\nabla^{\langle r}u^{r\rangle}, \label{STR}\\
\nabla^{\langle\varphi}u^{\tau\rangle}u_\tau + \nabla^{\langle\varphi}u^{r\rangle}u_r &= 0 
~~\Rightarrow~~ \nabla^{\langle\tau}u^{\varphi\rangle} = v\nabla^{\langle r}u^{\varphi\rangle}. \label{STP}
\end{align}

Therefore, from Eqs. (\ref{STT}) and (\ref{STR}), we see that
\begin{align}\label{STTF}
\nabla^{\langle\tau}u^{\tau\rangle} &= v\nabla^{\langle\tau}u^{r\rangle} = v^2\nabla^{\langle r}u^{r\rangle} \nonumber\\
&= \frac{1}{3}\left[\frac{(u^r)^2 u^{\tau}}{\tau} - \frac{(u^r)^3}{r(u^\tau)^2} + \frac{(u^{r})^5}{r(u^\tau)^2}\right].
\end{align}
Next, in order to verify our algebra, we confirm that the viscous 
stress tensor is traceless, i.e., $g_{\mu\nu}\nabla^{\langle\mu} 
u^{\nu\rangle}=0$. Using Eqs. (\ref {SPP}), (\ref{SRR}), (\ref{SEE}) 
and (\ref{STTF})
\begin{align}\label{check}
g_{\mu\nu}\nabla^{\langle\mu}u^{\nu\rangle} =&~ \nabla^{\langle\tau}u^{\tau\rangle} - 
\tau^2\nabla^{\langle\eta_s}u^{\eta_s\rangle} - \nabla^{\langle r}u^{r\rangle} - 
r^2\nabla^{\langle\varphi}u^{\varphi\rangle} \nonumber\\
=&~ \frac{1}{3}\left[\frac{(u^r)^2 u^{\tau}}{\tau} 
- \frac{(u^r)^3}{r(u^\tau)^2} + \frac{(u^{r})^5}{r(u^\tau)^2}\right]\nonumber\\
&- \frac{1}{3}\left[2\frac{u^{r}}{r} - 2\frac{u^{\tau}}{\tau} - \frac{(u^{r})^3}{r(u^\tau)^2}\right] \nonumber\\
&- \frac{1}{3}\left[\frac{(u^{\tau})^3}{\tau} - \frac{u^{r}}{r} + \frac{(u^{r})^3}{r}\right]\nonumber\\
&- \frac{1}{3}\left[\frac{u^{\tau}}{\tau} - \frac{u^{r}}{r} - \frac{(u^{r})^3}{r(u^\tau)^2}\right] \nonumber\\
=&~ 0.
\end{align}

For a particle at the space-time point $(\tau,\,\eta_s,\,r,\,\varphi)$ 
with the four momentum $p^\mu=(E,\,p^x,\,p^y,\,p^z)=(m_T\cosh y, 
\,p_T\cos\varphi_p,\,p_T\sin\varphi_p,\,m_T\sinh y)$, we get
\begin{align}\label{momenta}
p^\tau &= m_T\cosh(y-\eta_s) ~~\Rightarrow~~ p_\tau = m_T\cosh(y-\eta_s), \nonumber\\
\tau\, p^{\eta_s} &= m_T\sinh(y-\eta_s) ~~\Rightarrow~~ p_{\eta_s} = -\tau\, m_T\sinh(y-\eta_s), \nonumber\\
p^r &= p_T\cos(\varphi_p-\varphi) ~~\Rightarrow~~ p_r = -p_T\cos(\varphi_p-\varphi), \nonumber\\
r\,p^\varphi &= p_T\sin(\varphi_p-\varphi) ~~\Rightarrow~~ p_\varphi = -r\, p_T\sin(\varphi_p-\varphi).
\end{align}
The oriented freeze-out hyper-surface is $d\Sigma_\mu=(\tau 
d\eta_s\,rdr\,d\varphi,\,0,\,0,\,0)$, and therefore the integration 
measure is given by
\begin{equation}\label{intmes}
p^\mu d\Sigma_\mu = m_T\cosh(y-\eta_s)\,\tau d\eta_s\,rdr\,d\varphi.
\end{equation}
The viscous correction to the equilibrium distribution function is 
proportional to
\begin{align}\label{VisCorr}
p_\mu p_\nu\nabla^{\langle\mu}u^{\nu\rangle} =&~ p_\tau^2\nabla^{\langle\tau}u^{\tau\rangle} 
+ p_{\eta_s}^2\nabla^{\langle\eta_s}u^{\eta_s\rangle} 
+ p_r^2\nabla^{\langle r}u^{r\rangle} \nonumber\\
&+ p_\varphi^2\nabla^{\langle\varphi}u^{\varphi\rangle} 
\!+ 2p_\tau p_r\nabla^{\langle\tau}u^{r\rangle}
\!+ 2p_r p_\varphi\nabla^{\langle r}u^{\varphi\rangle} \nonumber\\
&+ 2p_\tau p_\varphi\nabla^{\langle\tau}u^{\varphi\rangle}.
\end{align}
The final form of $f=f_0+\delta f$ obtained from Eq.~(\ref{GradT}) 
using the above equation is required to evaluate the spectra given by
\begin{equation}\label{Final_Form}
\frac{d^{2}N}{d^{2}p_{T}\,dy} = \frac{1}{(2\pi)^3} 
\!\! \int_{0}^{R} \!\!\!\!\! r \, dr \!\! \int_{0}^{2\pi} \!\!\!\!\! d\varphi 
\!\! \int_{-\infty}^{\infty} \!\!\!\!\! \tau \, d\eta_s \, m_{T}\cosh(y -\eta_{s}) f.
\end{equation}



\begin{thebibliography}{99}

\bibitem{Adams:2005dq} 
  J.~Adams {\it et al.}  [STAR Collaboration],
  Nucl.\ Phys.\ A {\bf 757}, 102 (2005).

\bibitem{Adcox:2004mh} 
  K.~Adcox {\it et al.}  [PHENIX Collaboration],
  Nucl.\ Phys.\ A {\bf 757}, 184 (2005).

\bibitem{ALICE:2011ab} 
  K.~Aamodt {\it et al.}  [ALICE Collaboration],
  Phys.\ Rev.\ Lett.\  {\bf 107}, 032301 (2011).

\bibitem{ATLAS:2012at} 
  G.~Aad {\it et al.}  [ATLAS Collaboration],
  Phys.\ Rev.\ C {\bf 86}, 014907 (2012).
  
\bibitem{Chatrchyan:2013kba} 
  S.~Chatrchyan {\it et al.}  [CMS Collaboration],
  Phys.\ Rev.\ C {\bf 89}, 044906 (2014).
  
\bibitem{Romatschke:2007mq} 
  P.~Romatschke and U.~Romatschke,
  Phys.\ Rev.\ Lett.\  {\bf 99}, 172301 (2007).

\bibitem{Song:2007ux} 
  H.~Song and U.~W.~Heinz,
  Phys.\ Rev.\ C {\bf 77}, 064901 (2008).

\bibitem{Heinz:2013th} 
  U.~Heinz and R.~Snellings,
  Ann.\ Rev.\ Nucl.\ Part.\ Sci.\  {\bf 63}, 123 (2013).

\bibitem{Gale:2013da}
  C.~Gale, S.~Jeon and B.~Schenke,
  Int.\ J.\ Mod.\ Phys.\ A {\bf 28}, 1340011 (2013).

\bibitem{Luzum:2008cw}
  M.~Luzum and P.~Romatschke,
  Phys.\ Rev.\  C {\bf 78}, 034915 (2008).
  [Erratum {\it ibid.} C {\bf 79}, 039903(E) (2009)].

\bibitem{Song:2007fn}
  H.~Song and U.~Heinz,
  Phys.\ Lett.\  {\bf B658}, 279 (2008);
  Phys.\ Rev.\ C {\bf 78}, 024902 (2008);
  
\bibitem{Steinheimer:2007iy}
 J.~Steinheimer, M.~Bleicher, H.~Petersen, S.~Schramm, H.~Stocker and D.~Zschiesche,
 Phys.\ Rev.\ C {\bf 77}, 034901 (2008).

\bibitem{Dusling:2007gi}
  K.~Dusling and D.~Teaney,
  Phys. Rev. C {\bf 77}, 034905 (2008).

\bibitem{Molnar:2008xj}
  D.~Molnar and P.~Huovinen,
  J.\ Phys.\ G {\bf 35}, 104125 (2008).

\bibitem{Bozek:2009dw}
  P.~Bozek,
  Phys.\ Rev.\  C {\bf 81}, 034909 (2010).

\bibitem{Chaudhuri:2009hj}
  A.~K.~Chaudhuri,
  J.\ Phys.\ G {\bf 37}, 075011 (2010).

\bibitem{Xu:2007jv}
  Z.~Xu, C.~Greiner, and H.~St\"ocker,
  Phys.\ Rev.\ Lett.\  {\bf 101}, 082302 (2008).

\bibitem{Holopainen:2010gz} 
  H.~Holopainen, H.~Niemi and K.~J.~Eskola,
  Phys.\ Rev.\ C {\bf 83}, 034901 (2011).
  
\bibitem{Qiu:2011iv} 
  Z.~Qiu and U.~W.~Heinz,
  Phys.\ Rev.\ C {\bf 84}, 024911 (2011).
  
\bibitem{Song:2010mg} 
  H.~Song, S.~A.~Bass, U.~Heinz, T.~Hirano and C.~Shen,
  Phys.\ Rev.\ Lett.\  {\bf 106}, 192301 (2011)
  [Erratum-ibid.\  {\bf 109}, 139904 (2012)].

\bibitem{Schenke:2010rr} 
  B.~Schenke, S.~Jeon and C.~Gale,
  Phys.\ Rev.\ Lett.\  {\bf 106}, 042301 (2011);
  Phys.\ Rev.\ C {\bf 85}, 024901 (2012).

\bibitem{Qiu:2011hf} 
  Z.~Qiu, C.~Shen and U.~Heinz,
  Phys.\ Lett.\ B {\bf 707}, 151 (2012).

\bibitem{Bhalerao:2015iya} 
  R.~S.~Bhalerao, A.~Jaiswal and S.~Pal,
  Phys.\ Rev.\ C {\bf 92}, 014903 (2015).

\bibitem{Schnedermann:1993ws} 
  E.~Schnedermann, J.~Sollfrank and U.~W.~Heinz,
  Phys.\ Rev.\ C {\bf 48}, 2462 (1993).

\bibitem{Bjorken:1982qr} 
  J.~D.~Bjorken,
  Phys.\ Rev.\ D {\bf 27}, 140 (1983).

\bibitem{Huovinen:2001cy} 
  P.~Huovinen, P.~F.~Kolb, U.~W.~Heinz, P.~V.~Ruuskanen and S.~A.~Voloshin,
  Phys.\ Lett.\ B {\bf 503}, 58 (2001).

\bibitem{Adler:2001nb} 
  C.~Adler {\it et al.}  [STAR Collaboration],
  Phys.\ Rev.\ Lett.\  {\bf 87}, 182301 (2001).

\bibitem{Teaney:2003kp} 
  D.~Teaney,
  Phys.\ Rev.\ C {\bf 68}, 034913 (2003).

\bibitem{Tang:2008ud} 
  Z.~Tang, Y.~Xu, L.~Ruan, G.~van Buren, F.~Wang and Z.~Xu,
  Phys.\ Rev.\ C {\bf 79}, 051901 (2009).

\bibitem{Sun:2014rda} 
  X.~Sun, H.~Masui, A.~M.~Poskanzer and A.~Schmah,
  Phys.\ Rev.\ C {\bf 91}, 024903 (2015).

\bibitem{Teaney:2001av} 
  D.~Teaney, J.~Lauret and E.~V.~Shuryak,
  nucl-th/0110037.

\bibitem{Habich:2014jna} 
  M.~Habich, J.~L.~Nagle and P.~Romatschke,
  Eur.\ Phys.\ J.\ C {\bf 75}, 15 (2015).

\bibitem{Cooper:1974mv} 
  F.~Cooper and G.~Frye,
  Phys.\ Rev.\ D {\bf 10}, 186 (1974).

\bibitem{Grad}
H. Grad, Comm. Pure Appl. Math. {\bf 2}, 331 (1949).

\bibitem{Romatschke:2009im} 
  P.~Romatschke,
  Int.\ J.\ Mod.\ Phys.\ E {\bf 19}, 1 (2010).
  
\bibitem{Bhalerao:2013pza} 
  R.~S.~Bhalerao, A.~Jaiswal, S.~Pal and V.~Sreekanth,
  Phys.\ Rev.\ C {\bf 89}, 054903 (2014).
  
\bibitem{Staig:2010pn} 
  P.~Staig and E.~Shuryak,
  Phys.\ Rev.\ C {\bf 84}, 034908 (2011).
  
\bibitem{Shuryak:2013ke} 
  E.~Shuryak and I.~Zahed,
  Phys.\ Rev.\ C {\bf 88}, 044915 (2013).

\bibitem{Kolb:2001qz} 
  P.~F.~Kolb, U.~W.~Heinz, P.~Huovinen, K.~J.~Eskola and K.~Tuominen,
  Nucl.\ Phys.\ A {\bf 696}, 197 (2001).

\bibitem{Roy:2011xt} 
  V.~Roy and A.~K.~Chaudhuri,
  Phys.\ Lett.\ B {\bf 703}, 313 (2011).

\bibitem{Abelev:2013vea} 
  B.~Abelev {\it et al.}  [ALICE Collaboration],
  Phys.\ Rev.\ C {\bf 88}, 044910 (2013).

\bibitem{Retinskaya:2013gca} 
  E.~Retinskaya, M.~Luzum and J.~Y.~Ollitrault,
  Phys.\ Rev.\ C {\bf 89}, 014902 (2014).

\bibitem{Niemi:2012aj} 
  H.~Niemi, G.~S.~Denicol, H.~Holopainen and P.~Huovinen,
  Phys.\ Rev.\ C {\bf 87}, 054901 (2013).

\bibitem{Yan:2014nsa} 
  L.~Yan, J.~Y.~Ollitrault and A.~M.~Poskanzer,
  Phys.\ Lett.\ B {\bf 742}, 290 (2015).

\end{thebibliography}
\end{document}